\documentclass[aps,amssymb,amsmath, twocolumn, pra, superscriptaddress]{revtex4}
\usepackage{graphicx}
\usepackage{epsfig}
\usepackage{dcolumn}
\usepackage[up]{subfigure}
\usepackage{float}
\usepackage{dsfont}	
\usepackage{relsize}

\usepackage{epstopdf}
\newcommand\norm[1]{\left\lVert#1\right\rVert}
\usepackage{float}
\usepackage{ragged2e}
\usepackage{braket}
\usepackage[font=small,labelfont=bf,justification=raggedright, format=plain]{caption}
\usepackage{setspace}
\usepackage[colorlinks, citecolor = magenta]{hyperref}
\makeatletter
\begin{document}
\title{Quantum circuits for the realization of equivalent forms of one-dimensional discrete-time quantum walks on near-term quantum hardware }
\author{Shivani Singh}
\affiliation{The Institute of Mathematical Sciences, C. I. T. Campus, Taramani, Chennai 600113, India}
\affiliation{Homi Bhabha National Institute, Training School Complex, Anushakti Nagar, Mumbai 400094, India}
\author{C. Huerta Alderete}
\affiliation{Joint Quantum Institute, Department of Physics and Joint Center for Quantum Information and Computer Science, University of Maryland, College Park, MD 20742, USA}
\affiliation{Instituto Nacional de Astrof\'{i}sica, \'{O}ptica y Electr\'{o}nica,\\ Calle Luis Enrique Erro No. 1, Sta. Ma. Tonantzintla, Puebla Codigo Postal 72840, Mexico}
\author{Radhakrishnan Balu}
\affiliation{U.S. Army Research Laboratory, Computational and Information Sciences Directorate, Adelphi, Maryland 20783, USA}
\affiliation{Department of Mathematics \& Norbert Wiener Center for Harmonic Analysis and Applications, University of Maryland, College Park, MD20742}
\author{Christopher Monroe} 
\affiliation{Joint Quantum Institute, Department of Physics and Joint Center for Quantum Information and Computer Science, University of Maryland, College Park, MD 20742, USA}
\author{Norbert M. Linke}
\affiliation{Joint Quantum Institute, Department of Physics and Joint Center for Quantum Information and Computer Science, University of Maryland, College Park, MD 20742, USA}
\author{C. M. Chandrashekar}
\affiliation{The Institute of Mathematical Sciences, C. I. T. Campus, Taramani, Chennai 600113, India}
\affiliation{Homi Bhabha National Institute, Training School Complex, Anushakti Nagar, Mumbai 400094, India}

\begin{abstract}
Quantum walks are a promising framework for developing quantum algorithms and quantum simulations.  They represent an important test case for the application of quantum computers. Here we present different forms of discrete-time quantum walks (DTQWs) and show their equivalence for physical realizations. Using an appropriate digital mapping of the position space on which a walker evolves to the multi-qubit states of a quantum processor, we present different configurations of quantum circuits for the implementation of DTQWs in one-dimensional position space. We provide example circuits for a five-qubit processor and address scalability to higher dimensions as well as larger quantum processors. 
\end{abstract}
\maketitle
\section{\label{sec1}Introduction}
There is great interest in developing quantum algorithms for potential speedups over conventional computers, and progress is being made in mapping such algorithms to current technology\,\cite{Pre18, Mon19}. Device architecture, qubit connectivity, gate fidelity, and qubit coherence time are metrics that define the trade-off in designing device-specific circuits. Quantum walks\,\cite{kempe2003quantum, venegas2012quantum}, exploiting quantum superposition of multiple paths, have played an important role in the development of a wide variety of quantum algorithms. Examples include algorithms for quantum search\,\cite{CC03, Amb03, NKW03, Amb07, MSS07}, graph isomorphism problems\,\cite{DW08, GFZ10, BW11}, ranking nodes in a network\,\cite{PM12, PMC13, LTR15, CMC19}, and quantum simulation at low- and high-energy scales\,\cite{STR06, CBS10, MBD13, Cha13, MBD14, CB15, AFF16, MP16, MMK17, MMC17}.

There are two main variants of quantum walks, the discrete-time quantum walk (DTQW)\,\cite{AAK01, TFM03} and the continuous-time quantum walk (CTQW)\,\cite{FG98, GW03}. The DTQW is defined on a Hilbert space comprising internal states of the single particle called coin space and position space, with the evolution being driven by a position shift operator controlled by a quantum coin operator. The CTQW is defined directly on the position Hilbert space, with the evolution being driven by the Hamiltonian of the system and adjacency matrix of the position space. In both variants, the probability distribution of the particle spreads quadratically faster in position space compared to the classical random walk\,\cite{GVR1958, feynman1986quantum, 10.2307/3214153, aharonov1993quantum, CC03}. 

Due to the Hilbert space configuration of DTQWs one can define many different forms of quantum coin operators and position shift operators that control the dynamics leading to variants such as the standard DTQW, directed DTQW\,\cite{hoyer2009faster, montanaro2005quantum, chandrashekar2014quantum}, split-step DTQW\,\cite{kitagawa2012observation, asboth2012symmetries, mallick2016dirac}, and the Szegedy walk\,\cite{szegedy2004quantum}.  These models have been successfully used to mimic different quantum phenomena such as Dirac cellular automata\,\cite{meyer1996quantum, perez2016asymptotic, mallick2016dirac}, strong and weak localizations\,\cite{chandrashekar2012disorder, joye2012dynamical},  topological phases\,\cite{obuse2011topological, kitagawa2010exploring}, and many more. 

Experimental implementations of quantum walks  have been reported in cold atoms\,\cite{perets2008realization, karski2009quantum}, NMR systems\,\cite{JHX03,ryan2005experimental}, and photonic systems\,\cite{schreiber2010photons, peruzzo2010quantum, broome2010discrete, XTA16, XXJ18}.
DTQW implementations are ideally suited for lattice-based quantum systems in which the lattice site represents the position space. The DTQW is realized on an ion-trap system by the mapping position space to the motional phase space\,\cite{schmitz2009quantum, zahringer2010realization}. However, implementation of quantum walks on the quantum circuit is crucial to explore the practical realm of their algorithmic applications. The quantum-circuit-based implementation of DTQWs was first performed on a multiqubit NMR system\,\cite{ryan2005experimental}. On any hardware, limitations in the qubit number and coherence time restrict the number of steps that can be implemented. For implementation of  a DTQW on a quantum circuit, one needs to map the position state to the multiqubit state. Protocols using one such mapping on an $(N+1)$-qutrit superconducting system to implement $N$ steps of a DTQW was  been reported\,\cite{zhou2019protocol}. Recently, an optimal form of the quantum circuit for the realization of a DTQW on a five-qubit ion-trap quantum processor was presented and used for digital simulation of Dirac cellular automata\,\cite{DLF16}. Here we present a complete theory beyond the optimal form of quantum circuits that was used for the realization of Dirac cellular automata.

In this paper, we review different forms of DTQWs and show their equivalence concerning physical implementation on a quantum circuit. We also present various forms of quantum circuits that can be realized on a five-qubit quantum processor for the implementation of one-dimensional DTQWs. The circuits provided are for two variants of the DTQW, a standard QW and a directed QW,  that can be used to realize other forms of the DTQW and Dirac cellular automata. They can be further scaled up and generalized to implement multiparticle DTQWs and DTQW-based algorithms. 

\begin{figure}[h!]
  \begin{center}
  \includegraphics[width=0.5\textwidth]{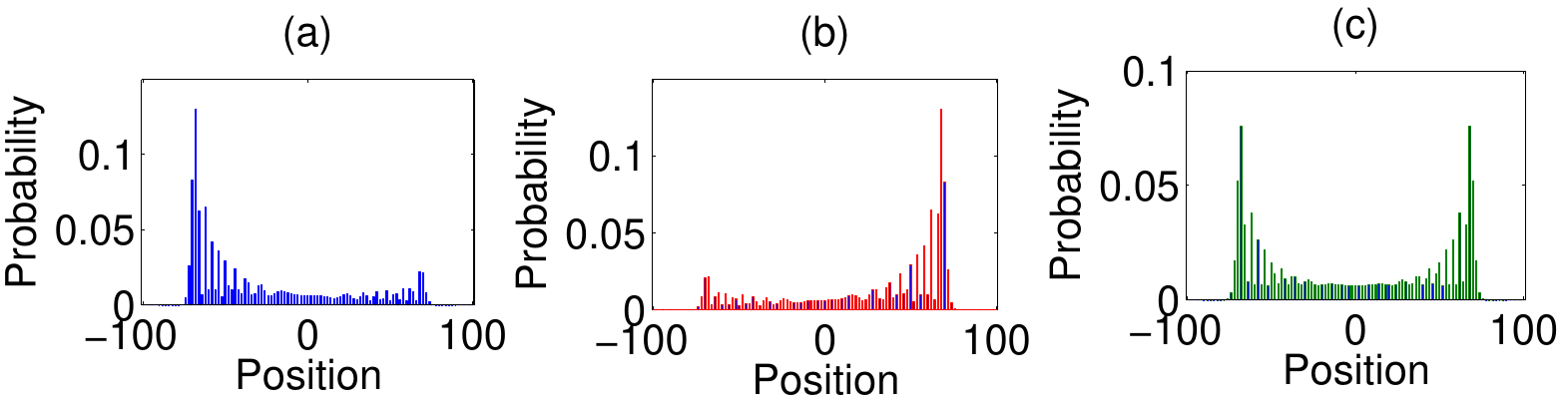}
  \end{center}
  \caption{Probability distribution after 100 time steps of a standard DTQW (SQW) for different initial states with the coin parameter $\theta = \pi/4$. Initial states are  $\ket{\Psi_{in}} = \ket{\uparrow} \otimes \ket{x = 0}$ for (a), $\ket{\Psi_{in}} = \ket{\downarrow} \otimes \ket{x = 0}$ for (b), and $\ket{\Psi_{in}} = \frac{1}{\sqrt{2}}(\ket{\uparrow} + \ket{\downarrow}) \otimes \ket{x = 0}$ for (c). Alternate sites will  have zero probability in a SQW irrespective of the initial state.} 
  \label{probSQW}
\end{figure}
\section{Variants of DTQWs and their equivalence }
\subsection{\label{sec2} Different forms of DTQWs}
DTQW is defined on the combination of particle (coin) and position Hilbert
space $\mathcal{H} =\mathcal{H}_c \otimes  \mathcal{H}_p$. Coin Hilbert space is defined by the particles' internal states $
\mathcal{H}_c  = \mbox{span} \{\ket{\uparrow},  \ket{\downarrow}\}$  and the one-dimensional position Hilbert space is spanned by $\mathcal{H}_p  = \mbox{span}
\{\ket{x}\}$, where $x  \in \mathbb{Z}$ represents the labels on the position states.
The generic initial state of the particle, $\ket{\psi}_c$, can be written as, 
\begin{equation}
\ket{\psi(\delta, \eta)}_c = \cos(\delta) \ket{\uparrow} + e^{-i\eta} \sin(\delta) \ket{\downarrow}.
\end{equation}
Each step of the walk evolves using a quantum coin operator acting on the particle space followed by a conditioned position shift operator acting on the entire Hilbert space. By modifying the coin and shift operators, different forms of DTQWs are achieved. The variants can have the same coin operations but have different shift operation. The coin operator with a single parameter is given by a rotation operator,
\begin{equation}
\hat{C}(\theta) = \begin{bmatrix}
   \cos(\theta) & -i \sin(\theta) \\
   -i \sin(\theta) & ~~ \cos(\theta) 
  \end{bmatrix} \otimes \mathcal{I}_l.
  \label{qcoin}
\end{equation}
Here $\mathcal{I}_l$ is the identity operator on the position space of length $l$.
\begin{figure}[h!]
  \centering
  \includegraphics[width=0.45\textwidth]{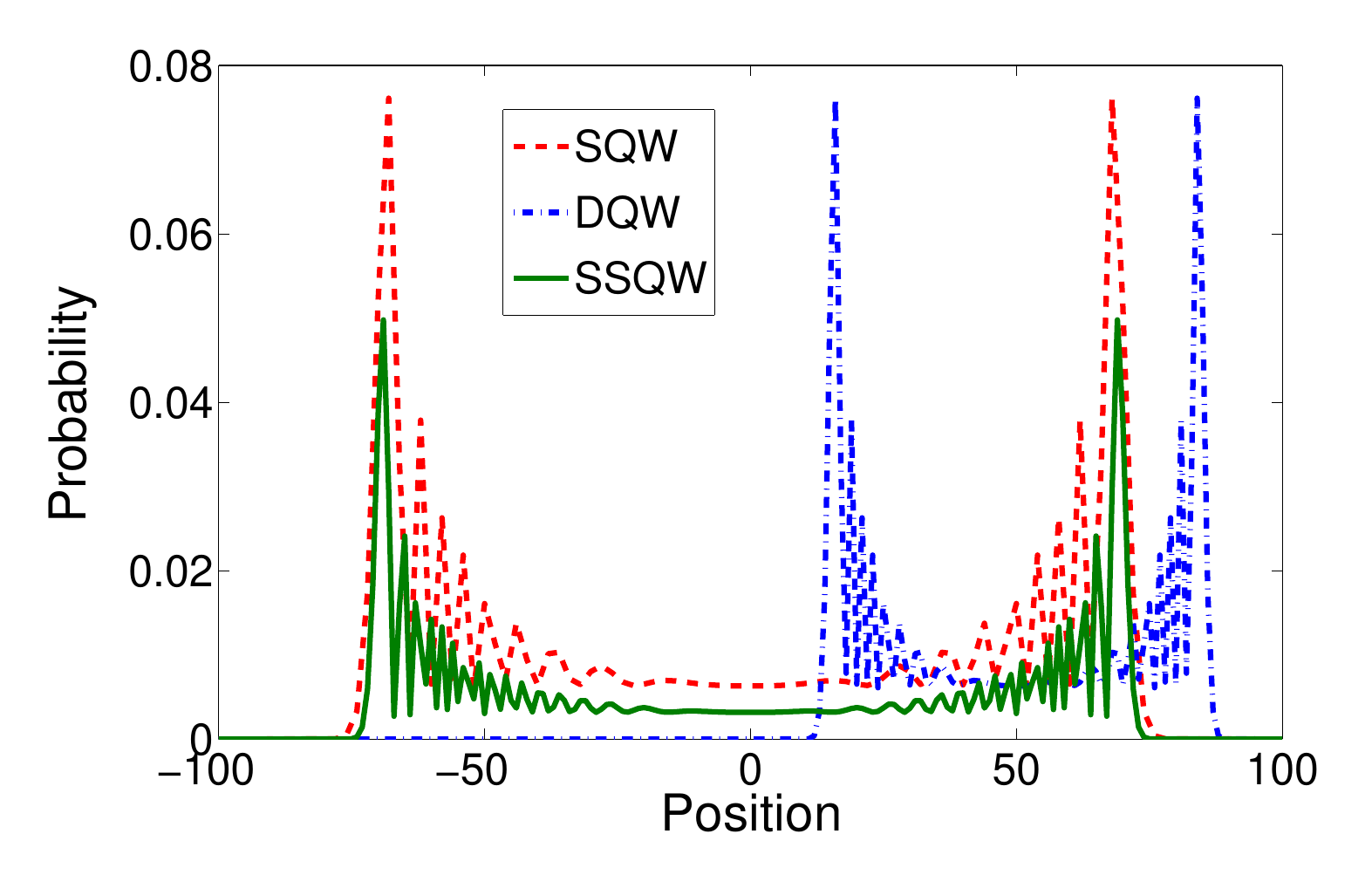}
  \caption{Probability distribution for standard DTQW (SQW), directed DTQW (DQW), and a split-step quantum walk (SSQW) with the coin parameter $\theta = \pi/4$ after 100-steps.  In the plot, the zero-probability values at alternate positions are discarded from the SQW. The spread in position space for the SQW and SSQW are identical, but the peak values of the distribution are different. The spread is different for SQW and DQW, but their peak values are identical. 
The initial state is $\ket{\Psi_{in}} = \frac{1}{\sqrt{2}}(\ket{\uparrow} + \ket{\downarrow}) \otimes \ket{x = 0}$ for all cases.} 
  \label{Prob_Standard}
\end{figure}

\noindent
{\it Standard DTQW  (SQW) :} Each step of SQW is realized by applying the operator  $\hat{W}=\hat{S}\hat{C}(\theta)$, where the coin operation for SQW is given by Eq.\,\eqref{qcoin} and the conditioned position shift operator  $\hat{S}$ is given by
\begin{equation} \label{Shift}
\hat{S}  = \sum_{x\in\mathbb{Z}} \bigg (\ket{\uparrow}\bra{\uparrow}
\otimes   \ket{x-1}\bra{x}+\ket{\downarrow}\bra{\downarrow} \otimes \ket{x+1}\bra{x}\bigg ).
\end{equation}
The state of the particle in extended position space after $t$ steps of a SQW is given by, 
\begin{equation}
\ket{\Psi(t)} = \hat{W}^t \bigg[\ket{\psi}_c \otimes \ket{x=0} \bigg ] =  \sum_{x=-t}^{t} \begin{bmatrix} \psi^{\uparrow}_{x, t} \\
 \psi^{\downarrow}_{x, t} \end{bmatrix}.
\end{equation}
The probability of finding the particle at position and time $(x, t)$ is
\begin{align}
P(x, t) = \norm{\psi^{\uparrow}_{x, t}}^2 + \norm{\psi^{\downarrow}_{x, t}}^2 .
\end{align}
Figure\,\ref{probSQW} shows the probability distribution of a SQW for different initial states for $\theta=\pi/4$. The symmetry of the probability distribution naturally depends on the particular  choice of the initial state of the walker. The symmetry and variance of the final distribution can also be affected by adding phases and thus taking advantage of the entire Bloch sphere for the coin operation in Eq.\,\eqref{qcoin}\,\cite{CSL08}.

\noindent
{\it Directed DTQW (DQW):} In one-dimensional position space each step of DQW is evolved  by applying the coin operation as given by Eq.\,\eqref{qcoin} on coin space followed by position shift operator  $\hat{S}_d$ of the form,
\begin{equation} \label{Shiftd}
\hat{S}_d  = \sum_{x\in\mathbb{Z}} \bigg ( \ket{\uparrow}\bra{\uparrow}
\otimes   \ket{x}\bra{x}+\ket{\downarrow}\bra{\downarrow} \otimes
\ket{x+1}\bra{x} \bigg ).
\end{equation}
The shift operator at time $t$ retains the particle at the existing position state or translates to the right conditioned on the internal state of the particle.  Each step of the walk is realized by applying the operator, $\hat{W}_d=\hat{S}_d\hat{C}(\theta)$. When the particle is in the superposition of the internal state, during each step of the walk, some amplitude of the particle  will simultaneously remain at the existing position state and translate to the right position state. In DQW, the probability amplitude is spread over the position space that is half of the position space compared to the spread of SQW.

\noindent
{\it Split-step DTQW (SSQW):}  In this variant, each step of the walk is a composition of two half-step evolutions, 
\begin{equation}
\label{DA}
\hat{W}_{ss} = \hat{S}_{+}\hat{C}(\theta) \hat{S}_{-}\hat{C}(\theta).
\end{equation}
The single parameter coin operator is again given by Eq.\,\eqref{qcoin} and the two shift operators have the forms
\begin{subequations}
\begin{align}
\hat{S}_{-} = \sum_{x\in\mathbb{Z}}\big ( \ket{\uparrow}\bra{\uparrow} \otimes \ket{x-1}\bra{x}+\ket{\downarrow}\bra{\downarrow} \otimes \ket{x}\bra{x} \big ), \\
\hat{S}_{+} =  \sum_{x\in\mathbb{Z}} \big ( \ket{\uparrow}\bra{\uparrow} \otimes  \ket{x}\bra{x}+\ket{\downarrow}\bra{\downarrow} \otimes \ket{x+1}\bra{x}\big ).
\end{align}
\label{shifta}
\end{subequations}
During each step of the SSQW, the particle remains at the same position and also moves to left and right positions  conditioned on the internal state of the particle. This leads to a probability distribution that is different from the SQW. In addition to that, a different value of $\theta$ can be used for each half step giving additional control over the dynamics and probability distribution.

In Fig.\,\ref{Prob_Standard} we show the probability distribution  over position space after 100 steps of a SQW, DQW and SSQW. The position space explored in the DQW is half the size compared to the SQW. The probability of finding the particle in each position space is non-zero for DQW when compared to SQW, in which the probability of finding particle at every alternate position is zero. Although the size of the position space is the same for both SSQW and SQW,  a nonzero probability of finding the particle at all positions is seen in the SSQW compared to the SQW, resulting in correspondingly lower peak values. 

\subsection{\label{sec1B} Equivalence of variants of discrete-time quantum walk }

Among the three forms of the walk presented above, the SSQW comprises both features, extended position states and non-zero probability at all positions. Therefore, one can consider SSQWs to be the most general form of a DTQW evolution. The state at any position $x$  and time $(t+1)$ after the operation of $\hat{W}_{ss}$ at time $t$ will be $\Psi_{x, t+1} = \psi^{\uparrow}_{x, t+1}  + \psi^{\downarrow}_{x, t+1}$, where
\begin{subequations}
\begin{align}
\psi^\uparrow_{x, t+1} &= \cos(\theta)[\cos(\theta)\psi^{\uparrow}_{x+1, t} - i \sin(\theta)\psi^{\downarrow}_{x+1, t}] \nonumber\\ 
&- i \sin(\theta) [-i \sin(\theta)\psi^{\uparrow}_{x, t}+ \cos(\theta)\psi^{\downarrow}_{x, t}],   \\
\psi^{\downarrow}_{x, t+1} &= -i  \sin(\theta)[\cos(\theta)\psi^{\uparrow}_{x, t} - i  \sin(\theta)\psi^{\downarrow}_{x, t}] \nonumber \\ 
&+  \cos(\theta) [-i \sin(\theta)\psi^{\uparrow}_{x-1, t}+ \cos(\theta)\psi^{\downarrow}_{x-1, t}].
\end{align}
\label{deqA}
\end{subequations}
In the description below, we show that the amplitudes of the walker positions in the different quantum walk variants are identical after relabeling of the position state, which establishes that they are all equivalent.

\noindent
{\it Equivalence of the SQW and SSQW:} If we evolve two steps of  the SQW we will arrive at a state that is identical to Eq.\,\eqref{deqA} with only a replacement of $\ket{x \pm 1}$ with $\ket{x \pm 2}$. Without loss of generality we can show that
\begin{align}\label{EQSSQW}
\hat{W}_{ss} &\equiv \hat{W}^2, \nonumber \\
\hat{S}_{+}\hat{C}(\theta)  \hat{S}_{-} \hat{C}(\theta) &\equiv \Big[\hat{S}\hat{C}(\theta)\Big]^2,
\end{align}
where
\begin{align}\label{SQWOpo}
\hat{W}_{ss} &= \hat{S}_{+}\hat{C}(\theta)  \hat{S}_{-} \hat{C}(\theta) \nonumber \\
= \Big[& \Big( \cos^{2}\theta\ket{\uparrow}\bra{\uparrow} - i \sin \theta \cos \theta  \ket{\uparrow}\bra{\downarrow}) \otimes \sum \ket{x-1}\bra{x} \nonumber \\
&+ (- i \sin \theta \cos \theta  \ket{\downarrow}\bra{\uparrow} - \sin^{2}\theta\ket{\downarrow}\bra{\downarrow} ) \otimes \sum \ket{x}\bra{x} \nonumber \\
&+ (- \sin^{2}\theta\ket{\uparrow}\bra{\uparrow} - i \sin \theta \cos \theta  \ket{\uparrow}\bra{\downarrow}  ) \otimes \sum \ket{x}\bra{x} \nonumber \\
+ (& - i \sin \theta \cos \theta  \ket{\uparrow}\bra{\downarrow} + \cos^{2}\theta\ket{\downarrow}\bra{\downarrow} ) \otimes \sum \ket{x+1}\bra{x} \Big. \Big]
\end{align}
and 
\begin{align}\label{SQWOp}
\hat{W}^2 &= \hat{S}\hat{C}(\theta)\hat{S}\hat{C}(\theta)  \nonumber \\
= \Big[ & \Big.(\cos^{2}\theta\ket{\uparrow}\bra{\uparrow} - i \sin \theta \cos \theta  \ket{\uparrow}\bra{\downarrow}) \otimes \sum \ket{x-2}\bra{x} \nonumber \\
& + (- i \sin \theta \cos \theta  \ket{\downarrow}\bra{\uparrow} - \sin^{2}\theta\ket{\downarrow}\bra{\downarrow} ) \otimes \sum \ket{x}\bra{x} \nonumber \\
& + (- \sin^{2}\theta\ket{\uparrow}\bra{\uparrow} - i \sin \theta \cos \theta  \ket{\uparrow}\bra{\downarrow}  ) \otimes \sum \ket{x}\bra{x} \nonumber \\
+ ( - & i \sin \theta \cos \theta  \ket{\uparrow}\bra{\downarrow} + \cos^{2}\theta\ket{\downarrow}\bra{\downarrow} ) \otimes \sum \ket{x+2}\bra{x} \Big. \Big].
\end{align}
The equivalence shown in Eq.\,\eqref{EQSSQW} can be established by mapping position space $|x \pm 2 \rangle$ to $|x \pm 1 \rangle$. The equivalence of Eqs.\,\eqref{SQWOpo} and \,\eqref{SQWOp}  can also be obtained by using a modified version of the shift operators $S^{\prime}_{-}$ and $S^{\prime}_{+}$ in which $|x\pm 1\rangle$  in $S_{-}$ and $S_{+}$ [Eq.\,\eqref{shifta}] is replaced with $|x\pm 2\rangle$. That is, 
\begin{align}\label{EQSSQWA}
\hat{W}^{\prime}_{ss}&  = \hat{S}^{\prime}_{+}\hat{C}(\theta)  \hat{S}^{\prime}_{-} \hat{C}(\theta)   = \Big[\hat{S}\hat{C}(\theta)\Big]^2.
\end{align}
Since the operator $\hat{W}_{ss} \equiv \hat{W}^{\prime}_{ss} = \Big[\hat{S}\hat{C}(\theta)\Big]^2$,   the
equivalence shown in Eq.\,\eqref{EQSSQW} can be established, and all three operators will execute an identical transformation when applied to any initial state. 
\begin{figure}[h!]
\centering
\includegraphics[width=0.46\textwidth]{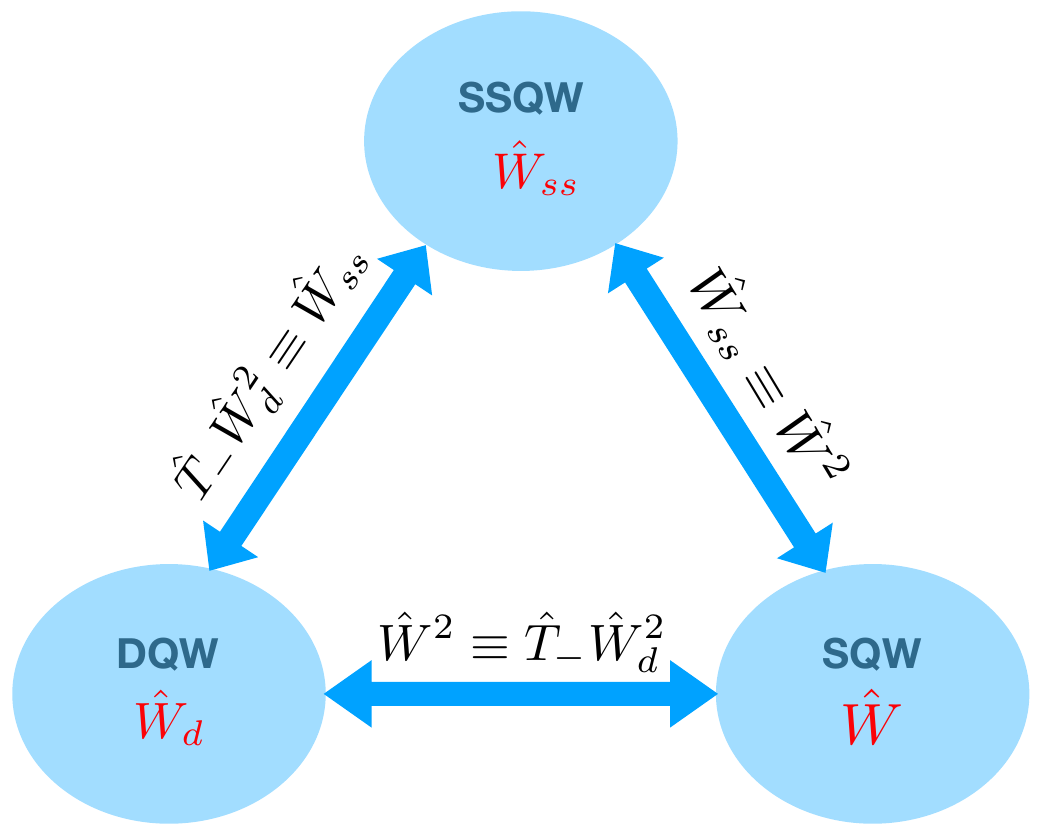}
\caption{Systematic presentation of the equivalence of the three forms of DTQWs.}
\label{Equivalence}
\end{figure}

\noindent
{\it Equivalence of the SQW and DQW :} Two SQW steps are equivalent to two DQW steps followed by a translation operator which executes a global shift on the position space. For the choice of shift operator that we have used, along with directed translation we can show that
\begin{align} \label{equivSD}
\hat{W}^2 &\equiv \hat{T}_{-} \hat{W_d}^2, \nonumber \\
\Big[\hat{S}\hat{C}(\theta)\Big]^2 &\equiv T_{-} \Big[ \hat{S_{d}}\hat{C}(\theta)\Big]^2,
\end{align}
where the forms of $\hat{C}(\theta)$, $\hat{S}$, and $\hat{S_d}$ are given in Eqs.\,\eqref{qcoin}, \eqref{Shift}, and \eqref{Shiftd}, respectively, and $\hat{T}_{-} = (\mathcal{I}_c \otimes \sum \ket{x-1}\bra{x})$. This can be explicitly shown by expanding the operators; $\hat{W}^2$ is given in Eq.\,\eqref{SQWOp}, and 
\begin{align} \label{TransD1}
\hat{W}_{TD} &= \hat{T}_{-}\hat{W_d}^2 \nonumber \\
&= \hat{T}_{-} \Big[ \hat{S}_{d}\hat{C}(\theta) \hat{S}_{d} \hat{C}(\theta)\Big] \nonumber \\
= \Big[& \Big.(\cos^{2}\theta\ket{\uparrow}\bra{\uparrow} - i \sin \theta \cos \theta  \ket{\uparrow}\bra{\downarrow}) \otimes \sum \ket{x-1}\bra{x} \nonumber \\
&+ (- i \sin \theta \cos \theta  \ket{\downarrow}\bra{\uparrow} - \sin^{2}\theta\ket{\downarrow}\bra{\downarrow} ) \otimes \sum \ket{x}\bra{x} \nonumber \\
&+ (- \sin^{2}\theta\ket{\uparrow}\bra{\uparrow} - i \sin \theta \cos \theta  \ket{\uparrow}\bra{\downarrow}  ) \otimes \sum \ket{x}\bra{x} \nonumber \\
+ ( - i & \sin \theta \cos \theta  \ket{\uparrow}\bra{\downarrow} + \cos^{2}\theta\ket{\downarrow}\bra{\downarrow} ) \otimes \sum \ket{x+1}\bra{x} \Big. \Big]. 
\end{align}
By replacing $x\pm 2$ with $x\pm 1$ in Eq.\,\eqref{SQWOp} we can show that $\hat{W}_{TD} \equiv \hat{W}^2$. Therefore, for all physical realizations mapping the position space of the walker onto multiqubit states of a quantum processor, one can ignore the alternate positions with zero probability in the SQW. A resulting probability distribution is equivalent to the translated DQW.

\noindent
{\it Equivalence of the SSQW and DQW}: A SSQW as described by the operator $\hat{W}_{ss}$ is equal to two DQW steps described by $\hat{W}_d$ followed by a global translation operator of the form $\hat{T}_{-} = (\mathcal{I}_c \otimes \sum \ket{x-1}\bra{x})$. The probability distribution of $2t$ time steps of the directed walk is the same as the probability distribution of $t$ steps of the split-step walk, i.e.,
\begin{align}\label{EqSSDQW}
\hat{W}_{ss} = \hat{T}_{-} \hat{W}_d^2,
\end{align} 
where $\hat{W}_{ss}$ and $\hat{W}_d$ are given in is given in Eqs. \eqref{DA} and  \eqref{Shiftd}, respectively. $\hat{T}_{-} \hat{W}_d^2$ is given in Eq. \eqref{TransD1}. 
Therefore, from Eqs.\,\eqref{EQSSQW}, \eqref{equivSD}, and \eqref{EqSSDQW} we get
\begin{align}
\hat{W}_{ss} = \hat{T}_{-}\hat{W}_{d}^2 \equiv \hat{W}^2.
\end{align}

This implies that $\psi^{\uparrow (\downarrow)}_{x \pm 1}=0$, i.e., the position with zero probability in the SQW. Thus, by discarding the positions with zero probability and relabelling values of position $x\pm 2$ as values of $x\pm 1$, the two-step SQW is equivalent to the SSQW\,\cite{kumar2018bounds}.

\begin{figure}[h!]
  \centering
  \includegraphics[width=0.48\textwidth]{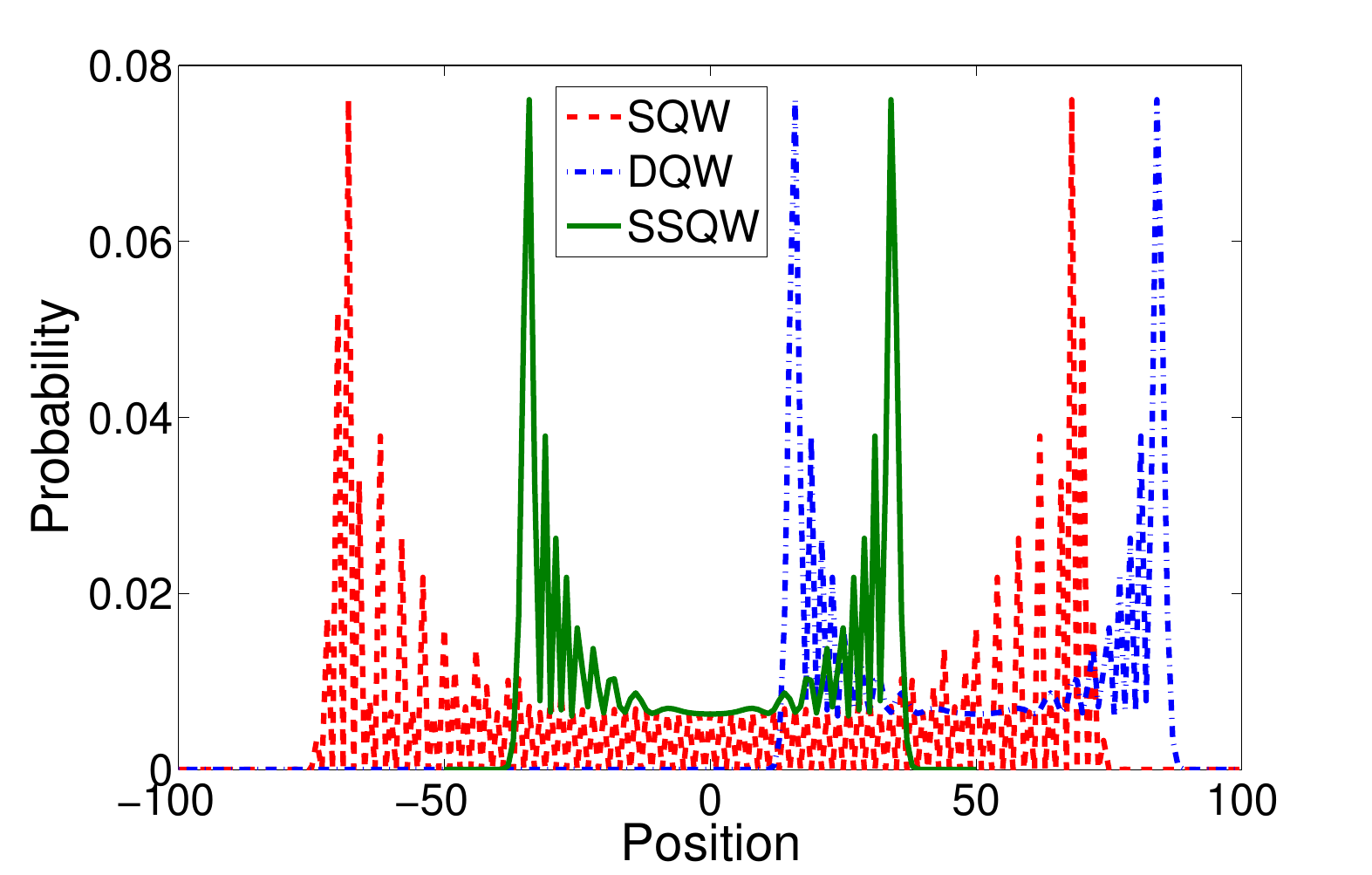}
  \caption{Equivalence of the probability distribution for different forms of DTQWs, i.e., the SQW and DQW for 100 steps and the SSQW for 50 steps, with the coin parameter $\theta = \pi/4$. Alternate sites of the SQW have zero probability, and thus 100 steps of SQW are equivalent to 50 time steps of the SSQW. The initial state is $\ket{\Psi_{in}} = \frac{1}{\sqrt{2}}(\ket{\uparrow} + \ket{\downarrow}) \otimes \ket{x = 0}$.} 
  \label{prob_Comparision}
\end{figure}

A schematic representation of the equivalence of all three forms of DTQWs is shown in  Fig.\,\ref{Equivalence}, while Fig.\,\ref{prob_Comparision} shows the probability distribution comparison for all three forms of DTQWs. The probability distribution of the SSQW is equivalent to half of the time evolution of the SQW and DQW. The probability values are the same for all three forms. Translation of the DQW in position space recovers the SSQW, and discarding of position space with zero probability in the SQW reduces its spread in position space and recovers the SSQW.

Therefore, a quantum circuit which can implement one form of the DTQW is sufficient to recover the exact probability distribution of the others by relabeling the position state associated with the multi-qubit state on the processor.  

\section{\label{sec3} Quantum circuit for implementing the DTQW}

To implement the DTQW on a quantum circuit in a one-dimensional position Hilbert space of size $2^N$, $(N+1)$ qubits are needed. Among $(N+1)$ qubits, one qubit acts as the coin, and the states of the remaining qubits are mapped to the position states in the DTQW. The basis for each qubit is characterized by its internal states, $\ket{0}$ and $\ket{1}$.  In principle, in $2^N$ position space,  $(2^{N-1}-1)$ steps of the SQW, and $(2^{N}-1)$-steps of the DQW  can be implemented.  %

\begin{figure}[h!]
\centering
\includegraphics[width=0.49\textwidth]{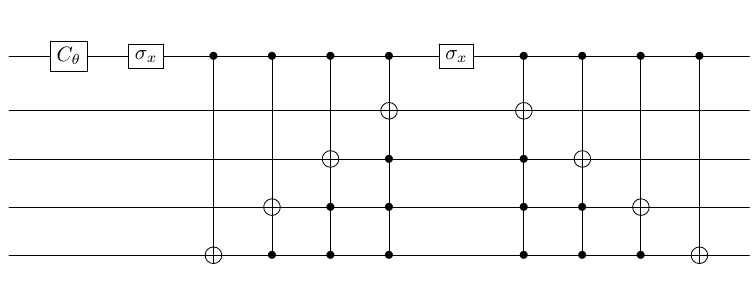}
\caption{ Generic quantum circuit to implement one step of the SQW on a five-qubit system for the mapping given in Table\,\ref{ConfigGen}. Repetition of this circuit will give us the SQW on the position Hilbert space with $16$ sites. The shift operation $S$ is performed using the increment and decrement circuit. }
\label{QW_5qubit}
\end{figure}

Each step of the SQW is evolved using a coin operation $C_{\theta}$ followed by the shift operation $S$ as given in Eq.\,\eqref{Shift}. Since $S$ acts on the position state and the mapping of the position to the qubit state is not unique, the composition of gates for the design of $S$ is also not unique. The coin operation $C_{\theta}$ can be carried out by using a single-qubit gate operation on the coin qubit, while the position shift operation $S$ can be subsequently applied with the help of multiqubit gates where the coin qubit acts as the control. For instance, Fig.\,\ref{QW_5qubit} presents a naive quantum circuit for a single step of the SQW on five-qubit quantum processor\,\cite{douglas2009efficient}.  The general form of this circuit depends on the mapping of the position state to the qubit states (see Table\,\ref{ConfigGen}). 
\begin{table}[H]
\caption{Mapping of the position state to the multiqubit state for the quantum circuit presented in Fig.\,\ref{QW_5qubit}. This multiqubit configuration identifies the even and odd position states in the system with the help of $|0\rangle$ and $|1\rangle$ as the state of the last qubit, respectively.}
\begin{tabular}{|m{11em} | m{0.005em}| m{11em}  | }
\hline
~~~$\ket{x = 0} \equiv \ket{0000}$  &&   \\
\hline
~~~$\ket{x = 1} \equiv \ket{0001}$ && ~~~$\ket{x = -1} \equiv \ket{1111}$\\
\hline 
~~~$\ket{x = 2} \equiv \ket{0010}$  &&  ~~~$\ket{x = -2} \equiv \ket{1110}$ \\
\hline
~~~$\ket{x = 3} \equiv \ket{0011}$ && ~~~$\ket{x = -3} \equiv \ket{1101}$\\
\hline
~~~$\ket{x = 4} \equiv \ket{0100}$  && ~~~$\ket{x = -4} \equiv \ket{1100}$\\
\hline
~~~$\ket{x = 5} \equiv \ket{0101}$ &&  ~~~$\ket{x = -5} \equiv \ket{1011}$\\
\hline 
~~~$\ket{x = 6} \equiv \ket{0110}$ && ~~~$\ket{x = -6} \equiv \ket{1010}$ \\
\hline
~~~$\ket{x = 7} \equiv \ket{0111}$ && ~~~$\ket{x = -7} \equiv \ket{1001}$ \\
\hline
\end{tabular}
\label{ConfigGen}
\end{table}
We note that the mapping of the position to the qubit state is not unique and the  quantum circuit can be simplified using different mapping. Here the odd (even)  position state is identified with the configuration of the last qubit $\ket{1}$ ($\ket{0}$). Repeating the circuit in Fig.\,\ref{QW_5qubit} will give seven steps of the SQW, but it can be scaled to $N$ qubits using an increment and decrement circuit.  However, for this mapping the gate size and gate counting per step of the SQW increase with the number of qubits.  
At this point, we have claimed that the quantum circuit complexity of the DTQW depends on the position space mapping. Therefore, we will now show that a mapping that takes the architecture of the quantum processor into account reduces the gate size and gate count. Additional reductions can be achieved by fixing the initial state of the walk. Here we present a quantum circuit on five-qubit system for the SQW and DQW that can be easily realized on present-day quantum processors, e.g., the five qubit programmable trapped-ion quantum computer\,\cite{DLF16} or IBM Quantum's five-qubit quantum computer\,\cite{FF2020, IBM}. 

 
\begin{table}[H]
\caption{Mapping of the position state to the multiqubit state for quantum circuits presented in Figs.\,\ref{SQW_General}, \ref{DQW_General}, \ref{CircuitSQW}, and \ref{CircuitDQW}. Here also the multiqubit configuration identifies even- and odd-numbered positions in the system with respect to the $|0\rangle$ and $|1\rangle$ states of the last qubit, respectively.}
\begin{tabular}{|m{11em} | m{0.005em}| m{11em}  | }
\hline
~~~$\ket{x = 0} \equiv \ket{0000}$  &&   \\
\hline
~~~$\ket{x = 1} \equiv \ket{0001}$ && ~~~$\ket{x = -1} \equiv \ket{0011}$\\
\hline 
~~~$\ket{x = 2} \equiv \ket{0110}$  &&  ~~~$\ket{x = -2} \equiv \ket{0010}$ \\
\hline
~~~$\ket{x = 3} \equiv \ket{0111}$ && ~~~$\ket{x = -3} \equiv \ket{0101}$\\
\hline
~~~$\ket{x = 4} \equiv \ket{1100}$  && ~~~$\ket{x = -4} \equiv \ket{0100}$\\
\hline
~~~$\ket{x = 5} \equiv \ket{1101}$ &&  ~~~$\ket{x = -5} \equiv \ket{1111}$\\
\hline 
~~~$\ket{x = 6} \equiv \ket{1010}$ && ~~~$\ket{x = -6} \equiv \ket{1110}$ \\
\hline
~~~$\ket{x = 7} \equiv \ket{1011}$ && ~~~$\ket{x = -7} \equiv \ket{1001}$ \\
\hline
\end{tabular}
\label{Congfig1}
\end{table}
\begin{figure}[h!]
\centering
\includegraphics[width=0.5\textwidth]{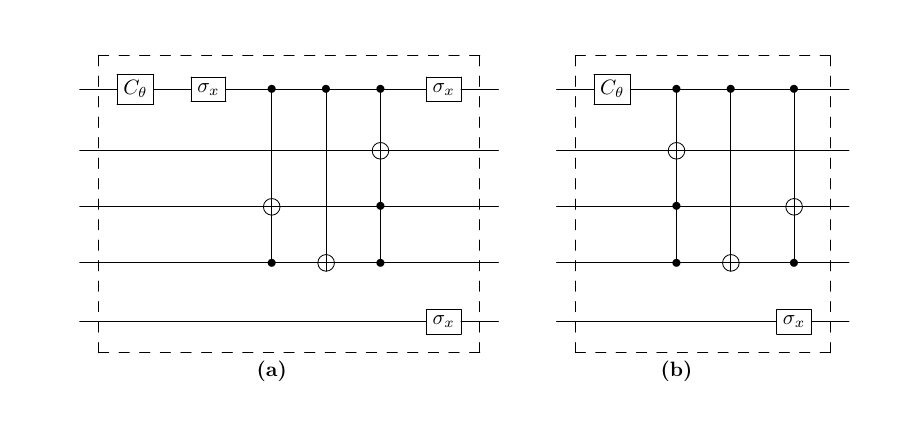}
\caption{ Generic quantum circuit for two steps of the SQW on a five-qubit system for the mapping given in Table\,\ref{Congfig1}. It can be used to implement up to seven steps of the SQW by alternating the circuits in (a) and (b). If the initial position state is even, the circuit in (a) is applied first, and if the initial position state is odd, the circuit in  (b) is applied first.}
\label{SQW_General}
\end{figure}
\begin{figure}[h!]
\centering
\includegraphics[width=0.45\textwidth]{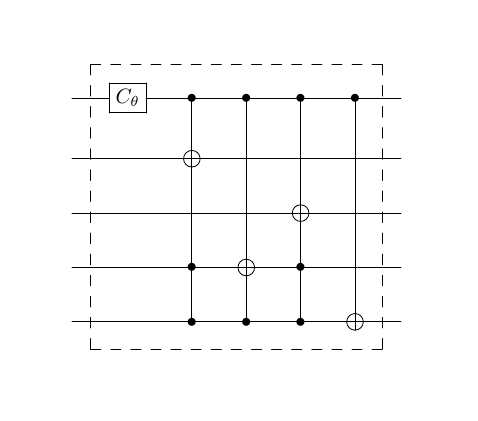}
\caption{Generic quantum circuit for a DQW on a five-qubit system for the mapping given in Table\,\ref{Congfig1}. Concatenation of this circuit will give the probability distribution of the DQW for up to $15$ steps.}
\label{DQW_General}
\end{figure}
\begin{table}[H]
\caption{Mapping of the position state to the multi-qubit state for quantum circuits presented in Figs.\,\ref{CircuitSQW2} and \,\ref{CircuitDQW2}.}
\begin{tabular}{|m{11em} | m{0.005em}| m{11em}  | }
\hline
~~~$\ket{x = 0} \equiv \ket{0000}$  &&   \\
\hline
~~~$\ket{x = 1} \equiv \ket{0001}$ && ~~~$\ket{x = -1} \equiv \ket{0111}$\\
\hline 
~~~$\ket{x = 2} \equiv \ket{0010}$  &&  ~~~$\ket{x = -2} \equiv \ket{0110}$ \\
\hline
~~~$\ket{x = 3} \equiv \ket{0011}$ && ~~~$\ket{x = -3} \equiv \ket{0101}$\\
\hline
~~~$\ket{x = 4} \equiv \ket{1100}$  && ~~~$\ket{x = -4} \equiv \ket{0100}$\\
\hline
~~~$\ket{x = 5} \equiv \ket{1101}$ &&  ~~~$\ket{x = -5} \equiv \ket{1011}$\\
\hline 
~~~$\ket{x = 6} \equiv \ket{1110}$ && ~~~$\ket{x = -6} \equiv \ket{1010}$ \\
\hline
~~~$\ket{x = 7} \equiv \ket{1111}$ && ~~~$\ket{x = -7} \equiv \ket{1001}$ \\
\hline
\end{tabular}
\label{Config2}
\end{table}


Figure\,\ref{SQW_General}, shows a quantum circuit for two steps of the SQW for the mapping presented in Table\,\ref{Congfig1}. Similar to the previous case, the state of the last qubit defines the even and odd positions. This allows us to keep the rest of the mapped qubits  identical for each pair of even and odd positions. For a generic initial position state $\ket{x}$ of the particle on a five-qubit system, the alternation of the circuits in Figs.\,\ref{SQW_General}(a) and \,\ref{SQW_General}(b) implements the seven steps of the SQW.   
If the initial position $\ket{x}$ is even (odd), the circuit in Figs.\,\ref{SQW_General}(a) [Figs.\,\ref{SQW_General}(b)] is applied first. When we compare the result to the quantum circuit in Fig.\,\ref{QW_5qubit} for a naive mapping, we see a significant decrease in the gate count and gate size for each step.  The gate count reduces to almost half for each step. 
Similarly, Fig.\,\ref{DQW_General} shows the quantum circuit for each step of the DQW for mapping presented in Table\,\ref{Congfig1} for any arbitrary initial position state $|x\rangle$, and with repeated application of this circuit, one can implement 15-steps of the DQW, in principle. 

In a comparison of the two quantum circuits for the SQW, one step of the naive mapping shown in Fig.\,\ref{QW_5qubit} has two  of each Toffoli-3 gate, Toffoli-2 gate, Toffoli gate, and controlled NOT (CNOT)  gate along with three single-qubit gates, while each step of the generic circuit shown in Fig.\,\ref{SQW_General} for mapping in Table\,\ref{Congfig1} has only one Toffoli-2 gate, one Toffoli gate, and one CNOT gate along with a few single-qubit gates.
Hence, this shows that for a smart position-state mapping, the gate count drops significantly, and hence, the circuit complexity decreases.   

Fixing the initial state of the walker helps to reduce the gate count in the quantum circuit and also reduces the circuit complexity. For example, for an initial state fixed to $\ket{0}\otimes \ket{0000} \equiv \ket{\uparrow} \otimes \ket{x=0}$,  the quantum circuits for the first seven steps of the SQW and DQW are shown in Figs.\,\ref{CircuitSQW} and \,\ref{CircuitDQW}, respectively. But for the implementation of the SSQW, two different shift operators are needed. The same results can be reconstructed from the equivalence relation between the SSQW and SQW, which will need two steps of the SQW to reproduce the results of the SSQW. Therefore, using the SQW and reconstructing the results of the corresponding SSQW from it are more efficient than the direct implementation of the SSQW.

We have also considered a different configuration of the position space mapping onto multiqubit states. As in Table\,\ref{Congfig1} the last qubit states, $|0\rangle$ and $|1\rangle$, are set to identify the even and odd position of the position state here too. The mapping given in Table\,\ref{Config2} and Figs.\,\ref{CircuitSQW2} and\,\ref{CircuitDQW2} shows the quantum circuits for the SQW and DQW for the mapping choices, respectively, which implements seven steps for the initial state $\ket{0} \otimes \ket{x=0}$. 
At alternate sites of the SQW we have zero probability, and our mapping allows the value of the last qubit to identify odd or even positions. Alternatively, the step number can be classically tracked in the quantum circuits shown in Figs.\,\ref{SQW_General}, \,\ref{CircuitSQW}, and \,\ref{CircuitSQW2} to reduce the number of $\sigma_x$ operations on the last qubit to zero or one.

Among the quantum circuits presented, the one given in Fig.\,\ref{CircuitSQW} is optimal for implementing the SQW. Table\,\ref{gatecount} gives a comparison of the number of gates in the optimized circuits in Fig.\,\ref{CircuitSQW},\,\ref{CircuitDQW},\,\ref{CircuitSQW2}, and\,\ref{CircuitDQW2}. 


\begin{table}[H]
\begin{center}
\caption{Gate count for mapping in Tables\,\ref{Congfig1}  and \ref{Config2} and for corresponding SQW and DQW circuits with the fixed initial state $\ket{0} \otimes \ket{0000}$ and after seven steps on a five-qubit quantum processor.}
\begin{tabular}{|m{4em}| m{7.5em} | m{7.5em} |}
\hline
 & SQW & DQW \\
\hline
 Table\,\ref{Congfig1}  & 22 single-qubit & 7 single-qubit \\
   & 7 two-qubit  & 7 two-qubit \\
   & 6 three-qubit & 6 three-qubit \\
   & 4 four-qubit & 10 four-qubit \\
\hline
 Table\,\ref{Config2} & 22 single-qubit & 7 single-qubit \\
   & 8 two-qubit & 6 two-qubit \\
   & 6 three-qubit & 6 three-qubit \\
   & 4 four-qubit & 8 four-qubit \\
\hline
\end{tabular}
\label{gatecount}
\end{center}
\end{table}

\begin{widetext}

\begin{figure}[H]
\includegraphics[width=\textwidth]{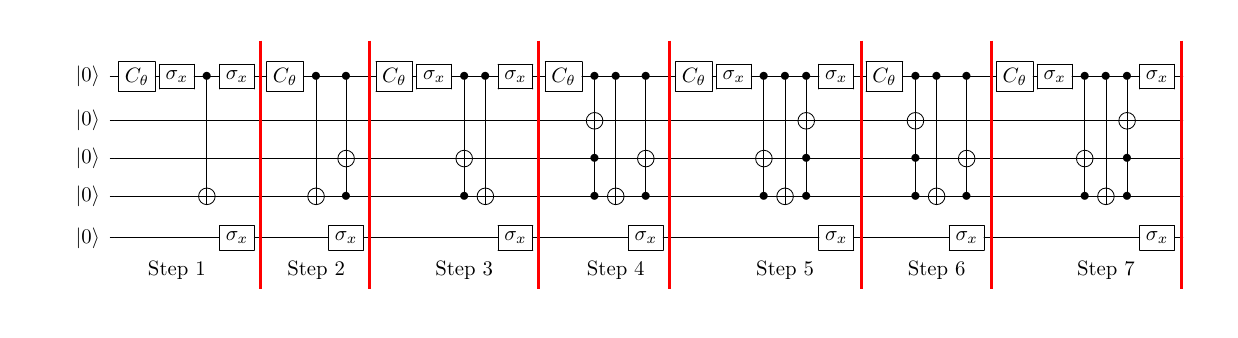}
\caption{Quantum circuit for the first seven steps of the SQW on a five-qubit system with the fixed initial state $\ket{\uparrow} \otimes \ket{x=0} \equiv \ket{0} \otimes \ket{0000}$ for the mapping given in Table\,\ref{Congfig1}. This circuit has a reduced gate count compared to the generic circuit shown in Fig.\,\ref{SQW_General}. We note that the sequence of $\sigma_x$ in the last qubit can be replaced by classically tracking the number of steps.}
\label{CircuitSQW}
\end{figure}
\begin{figure}[H]
\centering
\includegraphics[width=\textwidth]{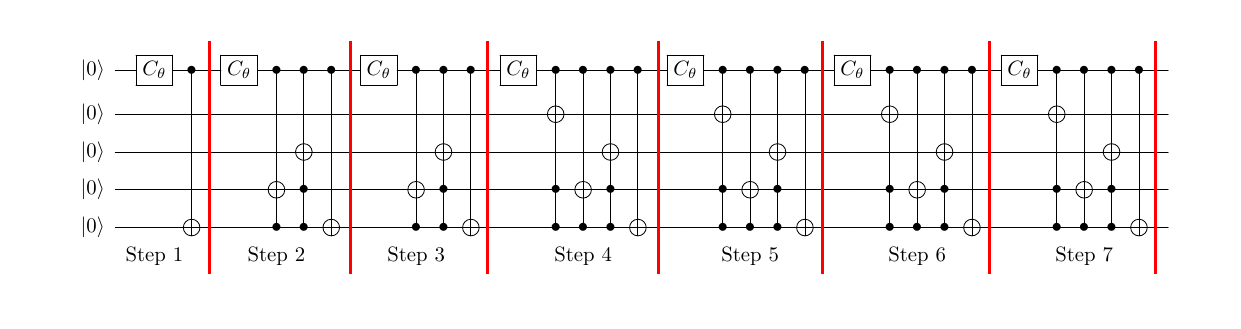}
\caption{Quantum circuit for the first seven steps of the DQW on a five-qubit system with the fixed initial state $\ket{\uparrow} \otimes \ket{x=0} \equiv \ket{0} \otimes \ket{0000}$ for the mapping given in  Table\,\ref{Congfig1}. It has a reduced gate count compared to the generic circuit shown in Fig.\,\ref{DQW_General}.}
\label{CircuitDQW}
\end{figure}
\begin{figure}[H]
\centering
\includegraphics[width=\textwidth]{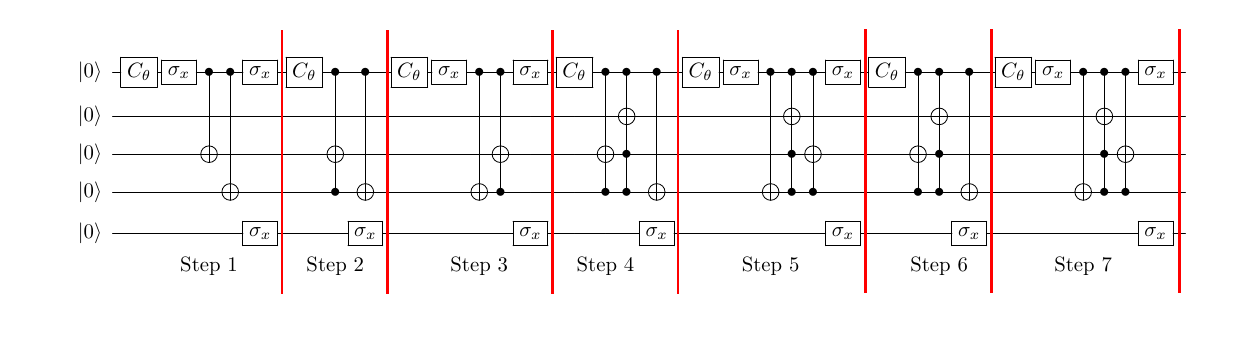}
\caption{Quantum circuit for the SQW for the first seven steps on a five-qubit system for the fixed initial state $\ket{\uparrow} \otimes \ket{x=0} \equiv \ket{0} \otimes \ket{0000}$ for the mapping given in Table\,\ref{Config2}. Here also, the sequence of $\sigma_x$ in the last qubit can be completely replaced by classicaly tracking the step number.}
\label{CircuitSQW2}
\end{figure}
\begin{figure}[H]
\centering
\includegraphics[width=\textwidth]{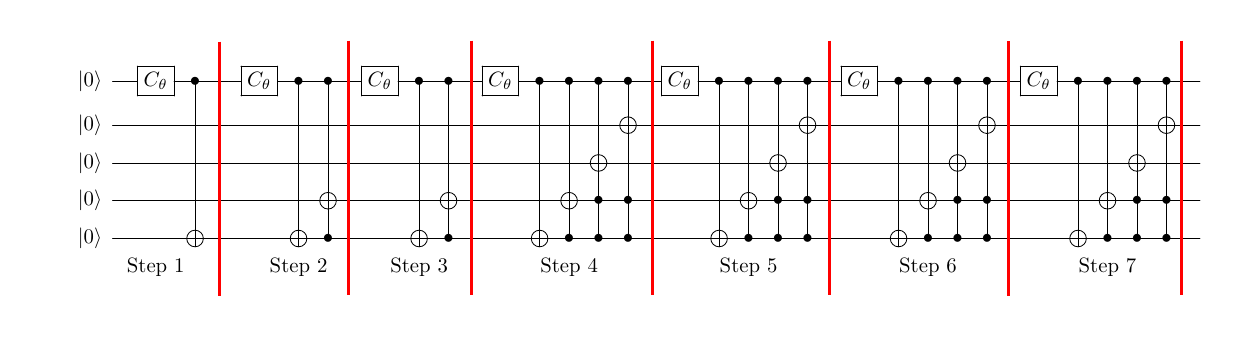}
\caption{Quantum circuit for the first seven steps of the DQW on a five-qubit system with the fixed initial state $\ket{\uparrow} \otimes \ket{x=0} \equiv \ket{0} \otimes \ket{0000}$ for the mapping given in Table\,\ref{Config2}.}
\label{CircuitDQW2}
\end{figure}

\end{widetext}
\section{\label{sec5} Discussion}

By digitally encoding the walker's position space onto the qubit state in various ways, we have shown different equivalent quantum walk circuits. The examples illustrate how the encoding methods and initial-state-dependent circuits can reduce the required gate depth (gate count) for implementing quantum walks.

The circuits can be scaled to implement more steps on a larger system using higher-order Toffoli gates. The implementation of $n$ steps of a SQW will need at least $[\log_{2}(n+1) + 2]$ qubits. Similarly, to implement $n$ steps of a DQW, at least $[\log_{2}(n+1) + 1]$ qubits are required. 

Recently, two different ways of expanding the increasing and decreasing parts of the generic quantum circuit shown in Fig.\ref{QW_5qubit} were explored in detail\,\cite{georgopoulos2021comparison}.  In one of the ways, the circuit complexity  is reduced by using the generalized controlled inversions method and in the other, the rotation operations around the basis states was used.  If the circuits are implemented on a device with a large number of qubits, then generalized controlled inversions would be a good option as they have less circuit depth due to the use of ancilla qubits or else the approach with rotation around basis state would be better. In our work the focus has been on reducing the circuit complexity by a careful choice of mapping of the qubit state to the position space and optimizing the circuit after choosing the initial state. Combining both these approaches may results in a further reduction of the circuit complexity, and that needs to be carefully explored in future works.

 DTQWs in two-dimensional position space\,\cite{FGB11, ChBu13} can also be implemented by scaling the scheme presented in this work with an appropriate mapping of qubit states with the nearest-neighbour position space in both dimensions.  This can be achieved on a device with access to a larger number of qubits by assigning an equal number of qubits to both dimensions in the two-dimensional position space and then by optimally mapping the qubit state to the position state.  All the circuits presented can be extended to implement two or more particle DTQWs by introducing two or more coin qubits into the system. In such cases, the control over the target or position qubit increases with the number of coin qubits. Another way of scaling  the scheme for the SQW on an $N$-qubit system, is to  fix one qubit for the coin as usual and another one to represent the $\pm$-sign for the positive and negative directions of the initial state $\ket{x}$, and the states of the rest of the $(N-2)$ qubits can be mapped to the position state. Now using the quantum adder circuit\,\cite{CT02}, the scheme can be extended to $N$ qubits, and a generalized quantum circuit for the quantum walk can be worked out.

One can also use ancilla qubits to reduce the circuit complexity. In the Appendix, we show a hybrid circuit with the help ancilla qubits for the DQW. The DQW can be implemented with the help of a CNOT gate, and interference in the walk can be included with the help of a controlled-SWAP gate and an ancilla qubit just before the measurement. Figures\,\ref{DQWancilla1} and \ref{DQWancilla2} show the hybrid circuit for three steps and four steps of the DQW. Details are is given in the Appendix.

Therefore, with an appropriate choice of the quantum coin operation and the equivalence of variants of DTQW, any quantum algorithm based on the DTQW can be experimentally realized on a quantum computer. Dirac cellular automata can be recovered using the SSQW, which reproduces the dynamics of the Dirac equation in the continuum limit\,\cite{mallick2016dirac}. One such example of simulating Dirac cellular automata on an ion-trap processor using one of the various configurations of the circuits presented  was demonstrated recently \cite{CS2020}. 
With the appropriate use of a position-dependent coin operation and additional higher-order Toffoli gates in our circuits, other DTQW-based algorithms, such as spatial search, can also be implemented. \\

{\bf Acknowledgments}\\
\par 
C.H.A acknowledges financial support from CONACYT Doctoral Grant No. 455378. N.M.L acknowledges financial support from NSF Grant No. PHY-1430094 to the PFC@JQI. C.M.C acknowledges the support from the Department of Science and Technology, Government of India, under Ramanujan Fellowship Grant No. SB/S2/RJN-192/2014 and U.S. Army ITC-PAC Grant No. FA520919PA139.


\newpage
\begin{widetext}

\begin{center}
{\bf Appendix}
\end{center}
\section*{\label{AppA} DQW circuit with naive mapping }
A naive mapping can result in an inefficient quantum circuit. One example of this is given in Table\, \ref{Config3} and Fig.\,\ref{DQWCircuit2}. 
\begin{table}[H] 
\begin{center}
\caption{Mapping of the position state onto multiqubit states for the DQW circuit presented in Fig.\,\ref{DQWCircuit2}. }
\begin{tabular}{|m{11em} | m{0.005em}| m{11em}  | }
\hline
~~~$\ket{x = 0} \equiv \ket{0000}$ &&~~ \\
\hline
~~~$\ket{x = 1} \equiv \ket{1000}$   &&  ~~ $\ket{x = 4} \equiv \ket{1111}$ \\
\hline 
~~~$\ket{x = 2} \equiv \ket{1100}$   &&  ~~ $\ket{x = 5} \equiv \ket{0111}$ \\
\hline
~~~$\ket{x = 3} \equiv \ket{1110}$  &&  ~~ $\ket{x = 6} \equiv \ket{1011}$ \\
\hline
\end{tabular}
\label{Config3}
\end{center}
\end{table}
 \begin{figure}[h!]
\centering
\includegraphics[width=0.7\textwidth]{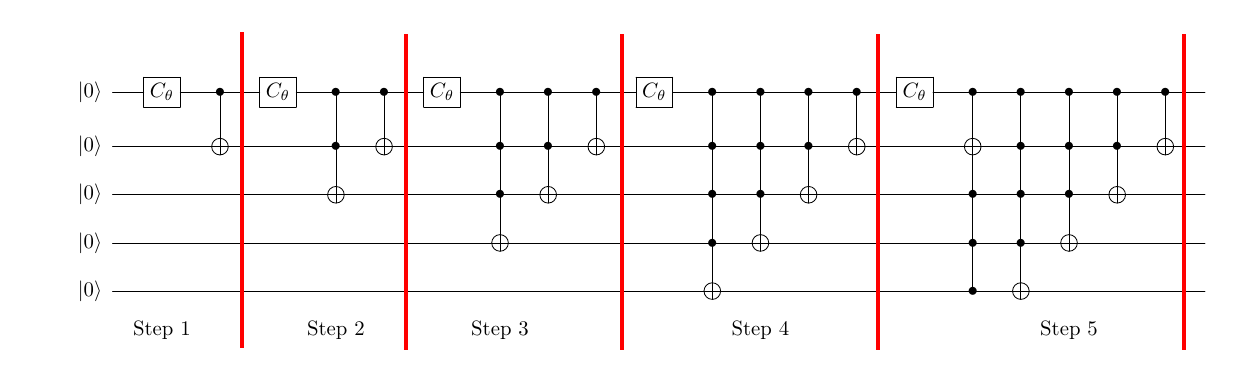}
\caption{Quantum circuit for the DQW for the first five steps with the fixed initial state $\ket{\Psi_{in}} = \ket{\uparrow} \otimes \ket{x=0} \equiv \ket{0} \otimes \ket{0000}$ for the naive mapping shown in Table \,\ref{Config3}. This circuit has a simple structure, but it consists of many additional higher-order Toffoli gates compared to the circuits shown in Sec.\,\ref{sec3} and the Appendix.}
\label{DQWCircuit2}
\end{figure}
The simplest quantum circuit for the mapping given above with fixed initial state, $\ket{0} \otimes |0000\rangle$ is shown in Fig.\,\ref{DQWCircuit2}. This circuit implements five steps of the DQW.  In the same system one can implement up to 15 steps since the available position states are $2^4=16$. This circuit looks straightforward to construct and scale but an actual implementation would require higher-order Toffoli gates even for a small number of steps and a fixed initial position, making it inefficient for near-term quantum processors.

\section*{\label{AppB} Simplified quantum circuit with an ancilla}
There has been a significant increase in the number of qubits available on platforms like trapped-ion and superconducting qubits\,\cite{landsman2019verified, zhang2017observation, bohnet2016quantum, yan2019strongly}. However, limited coherence time is still a hindrance to increasing the number of gates that can be implemented.  To make explicit use of the all available qubits, one has to develop low-depth quantum circuits. Here we will present quantum circuits with a reduced number of gates to implement DQWs at the cost of requiring additional ancilla qubits. But the given circuit is still inefficient as it will include only outputs with ancilla qubit state $\ket{0}$. In a system with access to more qubits, one can implement more steps of the DQW at the same circuit depth, but the efficiency decreases as the number of output states that can be included is when all the ancilla qubit state is $\ket{0}$ . 
\par 
For a five qubit system, we again use the first qubit to represent the coin  and the other four qubits to represent position space. The mapping is given in table\,\ref{Config4}. This is a classical circuit as it does not include the superposition or interference in the system directly. The output of the DQW and that of the quantum circuit in Fig.\,\ref{DQWcircuit3} is compared in table\,\ref{DQWOutput} for each step. To keep track of the contribution from each time evolution, we have introduced subscript to indicate different time steps. To turn this circuit into a DQW implementation, CNOT, Fredkin (controlled-Swap) gates and a Hadamard gate involving additional  ancilla qubits are applied before measurement as shown in Fig.\,\ref{DQWancilla1}. After measurement only selective outputs with ancilla qubit state $\ket{0}$ are included. 

\par 
\begin{table}[H] 
\begin{center}
\caption{Position state mapping used to construct the quantum circuit presented in Fig.\,\ref{DQWcircuit3}. This mapping requires ancilla qubits to induce interference by merging equivalent multi-qubit states.}
\begin{tabular}{|m{19.0em}  | }
\hline
~$\ket{x = 0} \equiv \ket{0000}$  \\
\hline
~$\ket{x = 1} \equiv \{ \ket{1000}, \ket{0100}, \ket{0010}, \ket{0001} \}$ \\
\hline 
~$\ket{x = 2} \equiv \{ \ket{1100}, \ket{1010}, \ket{1001}, $\\
$~~~~~~~~~~~~~~~ \ket{0110}, \ket{0101}, \ket{0011} \} $   \\
\hline
~$\ket{x = 3} \equiv \{ \ket{1110}, \ket{1101}, \ket{1011}, \ket{0111} \}$ \\
\hline
~$\ket{x = 4} \equiv \ket{1111}$  \\
\hline
\end{tabular}
\label{Config4}
\end{center}
\end{table}


\begin{table} [H]
\begin{center}
\caption{Output after each step of DQW and output of quantum circuit shown in Fig. \ref{DQWcircuit3} without the interference step provided by the ancilla circuit. Here $c_1, c_2,...$ represents the contribution of $\cos(\theta)$-term from the coin operation in the first, second ... time-evolutions and similarly $s_1, s_2,...$ represents the contribution of $\sin(\theta)$-term from the coin operation in the first, second ... time-evolutions, respectively in the circuit.}
\begin{tabular}{ | m{4em} | m{18em} | m{18em} | 
} 
\hline
$~~~$Steps &  Directed quantum walk output & Circuit (Fig.\,\ref{DQWcircuit3}) output without ancilla\\
\hline
$~~~~$0.    &     $\ket{0} \otimes \ket{x= 0} $      &      $ \ket{0} \otimes \ket{0000} $ \\ 
\hline 
$~~~~$1.    &     $c_1 \ket{0} \otimes \ket{x= 0} + s_1 \ket{1} \otimes \ket{x=1}$     & $c_1\ket{0} \otimes \ket{0000} + s_1\ket{1} \otimes \ket{1000} $ \\ 
\hline 
$~~~~$2.    &     $c_2 c_1 \ket{0} \otimes \ket{x= 0} + s_2 c_1 \ket{1} \otimes \ket{x=1} + s_2 s_1 \ket{0} \otimes \ket{x= 1} - c_2 s_1 \ket{1} \otimes \ket{x=2}$      &      $c_2 c_1\ket{0} \otimes \ket{0000} + s_2 c_1\ket{1} \otimes \ket{0100} + s_2 s_1 \ket{0} \otimes \ket{1000} - c_2 s_1 \ket{1} \otimes \ket{1100} $\\ 
\hline
$~~~~$3.    &     $c_3 c_2 c_1 \ket{0} \otimes \ket{x= 0} + s_3 c_2 c_1 \ket{1} \otimes \ket{x= 1} + s_3 s_2 c_1 \ket{0} \otimes \ket{x=1} - c_3 s_2 c_1 \ket{1} \otimes \ket{x=2} + c_3 s_2 s_1 \ket{0} \otimes \ket{x= 1} + s_3 s_2 s_1 \ket{1} \otimes \ket{x= 2} - s_3 c_2 s_1 \ket{0} \otimes \ket{x=2} + c_3 c_2 s_1 \ket{1} \otimes \ket{x=3} $     &     $c_3 c_2 c_1 \ket{0} \otimes \ket{0000} + s_3 c_2 c_1 \ket{1} \otimes \ket{0010} + s_3 s_2 c_1 \ket{0} \otimes \ket{0100} - c_3 s_2 c_1 \ket{1} \otimes \ket{0110} + c_3 s_2 s_1 \ket{0} \otimes \ket{1000} + s_3 s_2 s_1 \ket{1} \otimes \ket{1010} - s_3 c_2 s_1 \ket{0} \otimes \ket{1100} + c_3 c_2 s_1 \ket{1} \otimes \ket{1110}$\\ 
\hline
$~~~~$4.    &     $c_4 c_3 c_2 c_1 \ket{0} \otimes \ket{x= 0} + s_4 c_3 c_2 c_1 \ket{1} \otimes \ket{x= 1} + s_4 s_3 c_2 c_1 \ket{0} \otimes \ket{x= 1} - c_4 s_3 c_2 c_1 \ket{1} \otimes \ket{x= 2} + c_4 s_3 s_2 c_1 \ket{0} \otimes \ket{x=1} + s_4 s_3 s_2 c_1 \ket{1} \otimes \ket{x=2} - s_4 c_3 s_2 c_1 \ket{0} \otimes \ket{x=2} + c_4 c_3 s_2 c_1 \ket{1} \otimes \ket{x=3} + c_4 c_3 s_2 s_1 \ket{0} \otimes \ket{x= 1} + s_4 c_3 s_2 s_1 \ket{1} \otimes \ket{x= 2} + s_4 s_3 s_2 s_1 \ket{0} \otimes \ket{x= 2} - c_4 s_3 s_2 s_1 \ket{1} \otimes \ket{x= 3} - c_4 s_3 c_2 s_1 \ket{0} \otimes \ket{x=2} - s_4 s_3 c_2 s_1 \ket{1} \otimes \ket{x=3}  + s_4 c_3 c_2 s_1 \ket{0} \otimes \ket{x=3} - c_4 c_3 c_2 s_1 \ket{1} \otimes \ket{x=4}$        &       $c_4 c_3 c_2 c_1 \ket{0} \otimes \ket{0000} + s_4 c_3 c_2 c_1 \ket{1} \otimes \ket{0001} + s_4 s_3 c_2 c_1 \ket{0} \otimes \ket{0010} - c_4 s_3 c_2 c_1 \ket{1} \otimes \ket{0011} + c_4 s_3 s_2 c_1 \ket{0} \otimes \ket{0100} + s_4 s_3 s_2 c_1 \ket{1} \otimes \ket{0101} - s_4 c_3 s_2 c_1 \ket{0} \otimes \ket{0110} + c_4 c_3 s_2 c_1 \ket{1} \otimes \ket{0111} + c_4 c_3 s_2 s_1 \ket{0} \otimes \ket{1000} + s_4 c_3 s_2 s_1 \ket{1} \otimes \ket{1001} + s_4 s_3 s_2 s_1 \ket{0} \otimes \ket{1010} - c_4 s_3 s_2 s_1 \ket{1} \otimes \ket{1011} - c_4 s_3 c_2 s_1 \ket{0} \otimes \ket{1100} - s_4 s_3 c_2 s_1 \ket{1} \otimes \ket{1101}  + s_4 c_3 c_2 s_1 \ket{0} \otimes \ket{1110} - c_4 c_3 c_2 s_1 \ket{1} \otimes \ket{1111}$\\ 
\hline
\end{tabular}
\label{DQWOutput}
\end{center}
\end{table}
\begin{table} [H]
\begin{center}
\caption{Output after the three steps of a DQW using the quantum circuit shown in Fig.\,\ref{DQWcircuit3}
 and output of the quantum circuit with ancilla as shown in Fig.\,\ref{DQWancilla1} after the interference step. Here $c_1, c_2,...$ represents the contribution of $\cos(\theta)$-term from the coin operation in the first, second ... time-evolutions and similarly $s_1, s_2,...$ represents the contribution of $\sin(\theta)$-term from the coin operation in the first, second ... time-evolutions, respectively in the circuit.}
\begin{tabular}{ | m{4em} | m{18em} | m{18em} | 
} 
\hline
$~~~$Step & Circuit output without ancilla & Circuit output with ancilla \\
\hline
$~~~~$3.    &  $\Big( c_3 c_2 c_1 \ket{0} \otimes \ket{0000} + s_3 c_2 c_1 \ket{1} \otimes \ket{0010} + s_3 s_2 c_1 \ket{0} \otimes \ket{0100} - c_3 s_2 c_1 \ket{1} \otimes \ket{0110} + c_3 s_2 s_1 \ket{0} \otimes \ket{1000} + s_3 s_2 s_1 \ket{1} \otimes \ket{1010} - s_3 c_2 s_1 \ket{0} \otimes \ket{1100} + c_3 c_2 s_1 \ket{1} \otimes \ket{1110} \Big) \otimes \ket{0} $   &    $\Big( c_3 c_2 c_1 \ket{0} \otimes \ket{0000} \otimes \ket{0} + s_3 c_2 c_1 \ket{1} \otimes \ket{0010} \otimes \ket{0} + s_3 s_2 c_1 \ket{0} \otimes \ket{0100} \otimes \ket{0} + c_3 s_2 s_1 \ket{0} \otimes \ket{0100} \otimes \ket{1} + s_3 s_2 s_1 \ket{1} \otimes \ket{0110} \otimes \ket{1} - c_3 s_2 c_1 \ket{1} \otimes \ket{0110} \otimes \ket{0} - s_3 c_2 s_1 \ket{0} \otimes \ket{1100} \otimes \ket{1} + c_3 c_2 s_1 \ket{1} \otimes \ket{1110} \otimes \ket{1} \Big)$  \\ 
\hline
\end{tabular}
\label{Ancilla1}
\end{center}
\end{table}

\begin{table} [H]
\begin{center}
\caption{Output after the four steps of a DQW using the quantum circuit shown in Fig.\,\ref{DQWcircuit3} and output of the  quantum circuit with ancilla as shown in Fig.\,\ref{DQWancilla2} after the interference step. Here $c_1, c_2,...$ represents the contribution of $\cos(\theta)$-term from the coin operation in the first, second ... time-evolutions and similarly $s_1, s_2,...$ represents the contribution of $\sin(\theta)$-term from the coin operation in the first, second ... time-evolutions, respectively in the circuit.}
\begin{tabular}{ | m{4em} | m{18em} | m{18em} | 
} 
\hline
$~~$Step &  Circuit output without ancilla & Circuit Output with ancilla \\
\hline
 $~~~$ 4.    &      $\Big( c_4 c_3 c_2 c_1 \ket{0} \otimes \ket{0000} + s_4 c_3 c_2 c_1 \ket{1} \otimes \ket{0001} + s_4 s_3 c_2 c_1 \ket{0} \otimes \ket{0010} - c_4 s_3 c_2 c_1 \ket{1} \otimes \ket{0011} + c_4 s_3 s_2 c_1 \ket{0} \otimes \ket{0100} + s_4 s_3 s_2 c_1 \ket{1} \otimes \ket{0101} - s_4 c_3 s_2 c_1 \ket{0} \otimes \ket{0110} + c_4 c_3 s_2 c_1 \ket{1} \otimes \ket{0111} + c_4 c_3 s_2 s_1 \ket{0} \otimes \ket{1000} + s_4 c_3 s_2 s_1 \ket{1} \otimes \ket{1001} + s_4 s_3 s_2 s_1 \ket{0} \otimes \ket{1010} - c_4 s_3 s_2 s_1 \ket{1} \otimes \ket{1011} - c_4 s_3 c_2 s_1 \ket{0} \otimes \ket{1100} - s_4 s_3 c_2 s_1 \ket{1} \otimes \ket{1101}  + s_4 c_3 c_2 s_1 \ket{0} \otimes \ket{1110} - c_4 c_3 c_2 s_1 \ket{1} \otimes \ket{1111} \Big) \otimes \ket{000}$        &      
 $c_4 c_3 c_2 c_1 \ket{0} \otimes \ket{0000} \otimes \ket{000} + s_4 c_3 c_2 c_1 \ket{1} \otimes \ket{0001} \otimes \ket{000} + s_4 s_3 c_2 c_1 \ket{0} \otimes \ket{0100} \otimes \ket{001} + c_4 s_3 s_2 c_1 \ket{0} \otimes \ket{0100} \otimes \ket{000} + c_4 c_3 s_2 s_1 \ket{0} \otimes \ket{0100} \otimes \ket{100} + s_4 c_3 s_2 s_1 \ket{1} \otimes \ket{0101} \otimes \ket{100} - c_4 s_3 c_2 c_1 \ket{1} \otimes \ket{0101} \otimes \ket{001} + s_4 s_3 s_2 c_1 \ket{1} \otimes \ket{0101} \otimes \ket{000} + s_4 s_3 s_2 s_1 \ket{0} \otimes \ket{0110} \otimes \ket{101} - s_4 c_3 s_2 c_1 \ket{0} \otimes \ket{0110} \otimes \ket{001} - c_4 s_3 c_2 s_1 \ket{0} \otimes \ket{0110} \otimes \ket{010}+ c_4 c_3 s_2 c_1 \ket{1} \otimes \ket{0111} \otimes \ket{001} - c_4 s_3 s_2 s_1 \ket{1} \otimes \ket{0111} \otimes \ket{101} - s_4 s_3 c_2 s_1 \ket{1} \otimes \ket{0111} \otimes \ket{110} + s_4 c_3 c_2 s_1 \ket{0} \otimes \ket{1110} \otimes \ket{111} - c_4 c_3 c_2 s_1 \ket{1} \otimes \ket{1111} \otimes \ket{111}$\\ 
\hline
\end{tabular}
\label{ancilla2}
\end{center}
\end{table}


\begin{figure}[h!]
\centering
\includegraphics[width=0.5\textwidth]{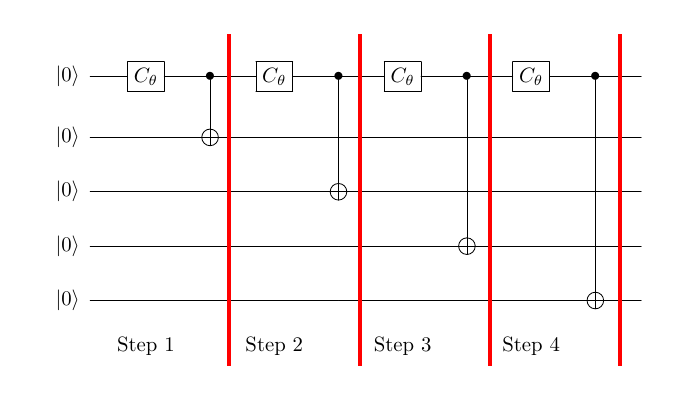}
\caption{Quantum circuit for DQW for first four steps without interference. Each step of this quantum circuit is given by a controlled-NOT gate because of the mapping chosen (see table\,\ref{Config4}). To include interference in the circuit, ancilla operations are needed before the measurement (see Fig.\,\ref{DQWancilla2}).}
\label{DQWcircuit3}
\end{figure}
After the first three steps, a single ancilla qubit introduces the equivalence of the states with two qubits in state $\ket{1}$ to position space at $\ket{x=2}$ as shown in Fig.\,\ref{DQWancilla1}. After the operation on ancilla qubit, the DQW distribution after 3 steps is recovered. Table\,\ref{Ancilla1}, shows the equivalence of the output of the third step of the DQW to the circuit output after the first three steps with an ancilla qubit operation before the Hadamard operation is performed on ancilla qubit.  
\begin{figure}[h!]
\centering
\includegraphics[width=0.5\textwidth]{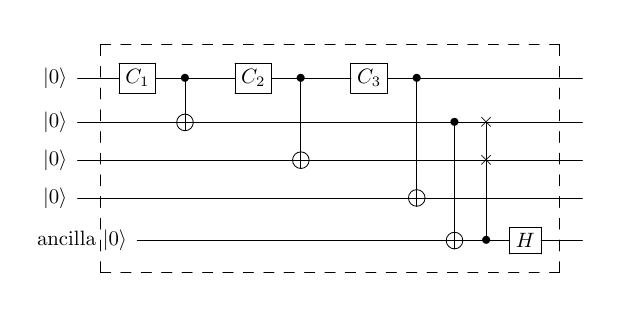}
\caption{Quantum circuit for DQW with the ancilla operation to include interference after first three steps. The ancilla qubit is left unobserved in the circuit. Here $C_1=C_2=C_3 = C_{\theta}$.}
\label{DQWancilla1}
\end{figure}
Similarly, to include interference after four steps, we need $3$ ancilla qubits as shown in Fig.\,\ref{DQWancilla2} and the output equivalence is shown in table\,\ref{ancilla2} before the Hadamard gate is performed on the ancilla qubit. The Hadamard operation helps in un-entangling the ancilla qubit with the real qubits in the circuit.
\begin{figure}[h!]
\centering
\includegraphics[width=0.50\textwidth]{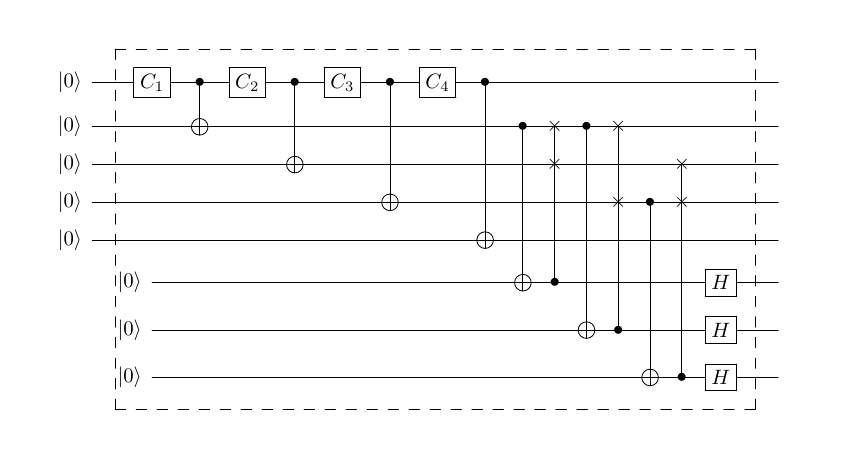}
\caption{Quantum circuit for DQW with ancilla operation to include interference after the first four steps. The ancilla qubits are left unobserved in the circuit. With a larger number of steps, the number of ancilla qubits also increases. Here $C_1=C_2=C_3 = C_4= C_{\theta}$.}
\label{DQWancilla2}
\end{figure}

\par 
 The number of ancilla qubits as well as  Fredkin (CSWAP) gates for the circuit in Fig.\,\ref{DQWcircuit3} increases as $^{n-1}C_{2}$ where, $n$ is the step number after which the measurement is done. The ancilla operation is needed only before the measurement. 
\par 
From Fig.\,\ref{DQWancilla2}, it can be seen that for the first four steps of DQW, the number of CNOT-gate required is 7 along with 3 Fredkin- gates. Each Fredkin- gate can be decomposed into 5 two-qubit gate\,\cite{SD96}. Therefore, total number of two-qubit gates required for first four steps of DQW using ancilla qubits are 22. Compared to this, if we look at the DQW circuit without ancilla, for the first four steps the number of CNOT gate required is 4,  number of Toffoli gate is 3, and number of controlled-Toffoli (CCCNOT) gate is 2 as can be seen in Figs.\,\ref{CircuitDQW} and \ref{CircuitDQW2}. Each Toffoli gate can be decomposed into 6 CNOT-gates and each CCCNOT gate can be further decomposed into two qubit gate and Toffoli gates\,\cite{NC00}.  The number of two-qubit gates required for first four steps of DQW without use of ancilla qubits  will be far more than the 22.  Therefore, a processor with access to large number of qubits with an possibility to leave anclilla qubit unobserved\,\cite{AES15, HPM19}  can be effective in reducing the gate counts to implement DTQW.

\end{widetext}

\end{document}